\documentclass[conference,compsoc]{IEEEtran}
\IEEEoverridecommandlockouts
\usepackage{cite}
\usepackage{amsmath,amssymb,amsfonts}
\usepackage{inconsolata}
\usepackage{algorithmic}
\usepackage{graphicx}
\usepackage{multirow}
\usepackage[hidelinks,colorlinks=true]{hyperref}
\usepackage{url}
\usepackage[noabbrev,capitalise]{cleveref}
\usepackage{textcomp}
\usepackage{xcolor}
\usepackage{tabularx}
\usepackage{booktabs}
\usepackage[noindentafter]{titlesec}
\usepackage[caption=false,font=normalsize,labelfont=sf,textfont=sf]{subfig}

\usepackage{enumitem}

\definecolor{myblue}{rgb}{0,0.3,0.6}
\hypersetup{linkcolor=myblue,urlcolor=myblue,citecolor=myblue,anchorcolor=myblue}
\crefformat{footnote}{#2\footnotemark[#1]#3}
\crefrangeformat{footnote}{#3\footnotemark[#1]#4--#5\footnotemark[#2]#6}
\crefmultiformat{footnote}{#2\footnotemark[#1]#3}{\textsuperscript{,}#2\footnotemark[#1]#3}{\textsuperscript{,}#2\footnotemark[#1]#3}{\textsuperscript{,}#2\footnotemark[#1]#3}
\setlist{parsep=0ex,topsep=0.5ex,itemsep=0ex,leftmargin=1.5em}

\newcolumntype{C}{>{\centering\arraybackslash}X}
\def\BibTeX{{\rm B\kern-.05em{\sc i\kern-.025em b}\kern-.08em
    T\kern-.1667em\lower.7ex\hbox{E}\kern-.125emX}}

\titleformat{\paragraph}[runin]{\normalfont\fontseries{bm}\selectfont}{\theparagraph}{1em}{}
\titlespacing{\paragraph}{0pt}{1ex}{1em}

\begin{document}

\urlstyle{tt}

\title{Equipping Pretrained Unconditional Music Transformers\\with Instrument and Genre Controls} 

\author{
    \IEEEauthorblockN{Weihan Xu\IEEEauthorrefmark{1} \quad Julian McAuley\IEEEauthorrefmark{2} \quad Shlomo Dubnov\IEEEauthorrefmark{2} \quad Hao-Wen Dong\IEEEauthorrefmark{2}}
    \IEEEauthorblockA{\IEEEauthorrefmark{1}Duke University \quad \IEEEauthorrefmark{2}University of California San Diego\\
    \hypersetup{linkcolor=black,urlcolor=black}
    \href{mailto:weihan.xu@duke.edu}{weihan.xu@duke.edu} \quad \href{mailto:jmcauley@ucsd.edu}{jmcauley@ucsd.edu} \quad \href{mailto:sdubnov@ucsd.edu}{sdubnov@ucsd.edu} \quad \href{mailto:hwdong@ucsd.edu}{hwdong@ucsd.edu}}
}

\maketitle

\begin{abstract}
The ``pretraining-and-finetuning'' paradigm has become a norm for training domain-specific models in natural language processing and computer vision. In this work, we aim to examine this paradigm for symbolic music generation through leveraging the largest ever symbolic music dataset sourced from the MuseScore forum. We first pretrain a large unconditional transformer model using 1.5 million songs. We then propose a simple technique to equip this pretrained unconditional music transformer model with instrument and genre controls by finetuning the model with additional control tokens. Our proposed representation offers improved high-level controllability and expressiveness against two existing representations. The experimental results show that the proposed model can successfully generate music with user-specified instruments and genre.

In a subjective listening test, the proposed model outperforms the pretrained baseline model in terms of coherence, harmony, arrangement and overall quality.
\end{abstract}


\setlength{\abovedisplayskip}{1ex}
\setlength{\belowdisplayskip}{1ex}
\setlength{\abovedisplayshortskip}{1ex}
\setlength{\belowdisplayshortskip}{1ex}

\section{Introduction}
AI-driven music generation empowers people 
without formal music training to create music \cite{briot 2017,AISONG}.
On one hand, while many music generation models (e.g., \cite{musegan,museformer}) offer the capability for unconditional music generation, these systems offer limited controllability, e.g., users cannot specify the genre and instrumentation of the generated music.
On the other hand, while some systems provide fine-grained controls for controllable music generation (e.g., \cite{FIGARO,popmag}), these controls are usually too complex for amateurs without a formal musical background.
In this work, we aim to find a middle ground by equipping a pretrained unconditional music generation system with high-level controls such as genre and instrumentation to improve its controllablilty while keeping its accessibility to nonprofessional users.
Various fine-grained control has been employed in music generation, including conditioning the model on preceding musical material, inpainting to generate music based on a partial subset of musical content, and requesting the model to fill in the gaps. For example, Music Sketchnet \cite{sketchnet} presented a framework designed to create the missing measures in incomplete monophonic musical compositions. This process was influenced by the surrounding context and can be enhanced by incorporating user-defined pitch and rhythm fragments if desired. PopMAG \cite{popmag} generated multitrack pop music accompaniment by conditioning on chord and melody. 
Additionally, users can also specify artist, genre and emotion information as conditions.
MuseCoco \cite{musecoco} extracted attributes such as genre and instruments from plain text to guide music generation. MuseCoco was trained on multiple MIDI datasets, including MMD and MetaMIDI datasets, whereas our model is trained on the larger MuseScore dataset. 
GetMusic \cite{getmusic} explored using a nonautoregressive model to generate music with any source-target track combinations. While GetMusic is trained on the MuseScore dataset, it lacks high-level controls over genre information. 
Multitrack Music Transformer (MMT) \cite{MMT} can generate multitrack music for a set of instruments specified by the user but it lacks control over genres. 

\begin{figure*}
    \footnotesize
    \centering
    \begin{tabular}{cc}
         \includegraphics[width=.48\linewidth]{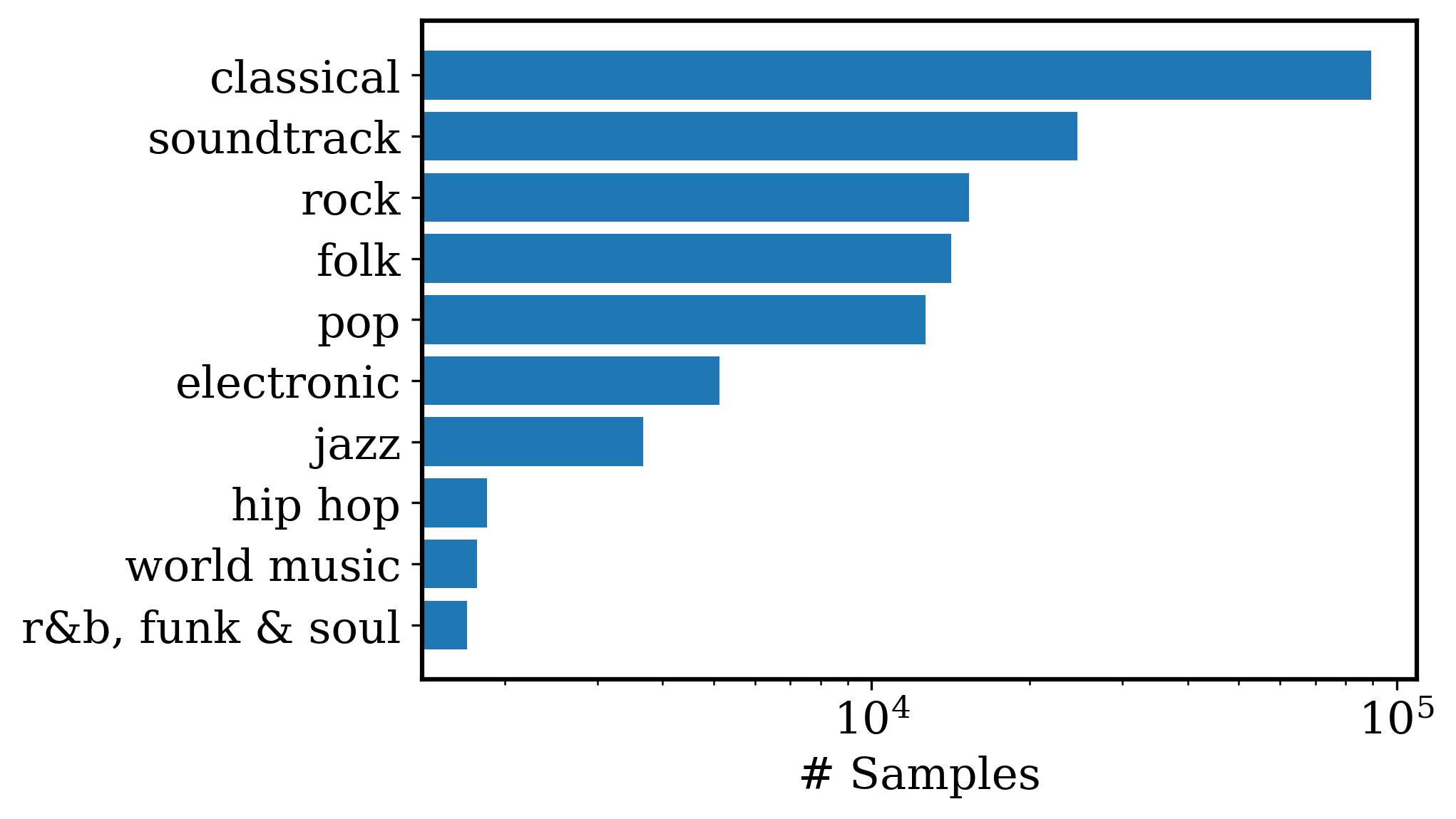} & \includegraphics[width=.47\linewidth]{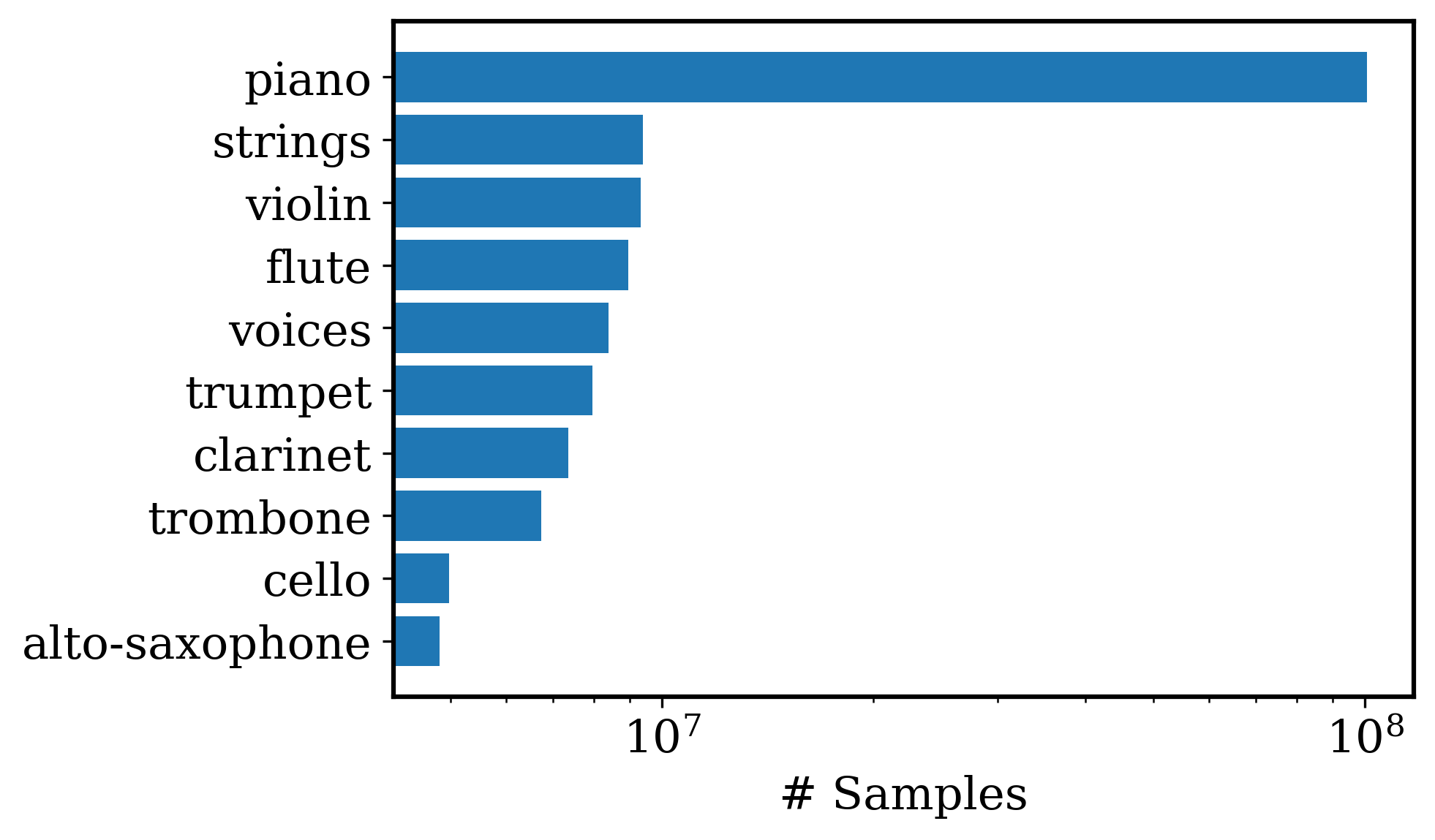}\\
         (a) Most common genres & (b) Most common instruments
    \end{tabular}
    \caption{Most common genres and instruments in the MuseScore dataset}
    \label{fig:condition}
\end{figure*}

Prior work has proposed various representations for multitrack music. For example, REMI
\cite{REMI} and REMI+\cite{FIGARO}, while encompassing a range of high-level musical information in a unified one-dimensional framework, do not allow for conditioning on genres and instruments. Moreover, Multitrack Music Transformer (MMT) \cite{MMT} proposes a multitrack music representation with multi-dimensional inputs and outputs to tackle memory complexity issues. Although MMT offers instrumentation controls, it is not as expressive as REMI+. Therefore, we aim to design a new representation that brings together the expressiveness of REMI+ with the high-level controls available in MMT.

In this work, we develop a method to augment a pretrained unconditional music transformer with instrument and genre controls by adding new control tokens to the model during the finetuning stage.
We train our proposed models on the MuseScore dataset, the largest-ever symbolic music dataset containing 1.5 million songs sourced from the MuseScore forum.\footnote{\label{fn:musescore}\url{https://musescore.com/}}
We evaluate the models through objective evaluation metrics and a subjective listening test.
Our experiment results show that our proposed models can effectively generate music adhere to user-specified instruments and genre. The proposed model outperforms the pretrained baseline model in terms of coherence, harmony and arrangement as well as overall quality in the subjective listening test.
Audio samples can be found on an anonymous demo website.\footnote{\url{https://goatlazy.github.io/MUSICAI/}\label{fn:demo}}

%==========================
\section{MuseScore Dataset}
%==========================
\label{AA}

We use a collection of 1.5 million sheet music files scraped from the MuseScore forum, an online forum where users can upload sheet music for existing songs and their own compositions.\cref{fn:musescore}
\cref{tab:common dataset} summarizes the differences between several common multitrack music dataset.
Moreover, we scraped the metadata from the MuseScore forum, which contains useful information such as genre, composer, rating, etc. There are 20 different genres and we use 64 different instruments in this dataset.
\cref{fig:condition} shows the statistics of most common genres and instruments in the MuseScore dataset.

In this work, we consider three subsets of the MuseScore dataset that offer different levels of information for training models with different controls:
\begin{itemize}
    \item
        \textbf{MuseScore-full} (1.5M): We extract information from the MuseScore dataset using MusPy package~\cite{muspy}, producing MusPy JSON files. From these files, we further extract notes information and form MuseScore-full dataset. This dataset is used to train the unconditional pretrained model.
    \item
        \textbf{MuseScore-metadata} (821K):
        We notice that half of the samples in MuseScore-full do not come with a metadata file. The MuseScore-metadata subset contains only files that have a matching metadata file. This dataset is used to train the instrument conditioned model.
    \item
        \textbf{MuseScore-genre} (175K): We take a subset of MuseScore-metadata by filtering out those songs without genre information and form MuseScore-genre dataset. This dataset is used to train the genre conditioned and the genre-instrument conditioned model.
\end{itemize}

\begin{table}
    \footnotesize
    \centering
    \caption{Large Datasets Commonly Used for Multitrack Music Generation}
    \label{tab:common dataset}
    \begin{tabular}{lccccc}
            \toprule
            Dataset             &\shortstack{Publicly\\available}  &\shortstack{Needs\\scraping} &Format &Songs\\
            \midrule
            LMD~\cite{raffle}           &\checkmark &$\times$   &MIDI &175K\\ 
            MetaMIDI~\cite{metamidi}    &\checkmark &$\times$   &MIDI &436K\\
            MMD~\cite{musecoco}         &$\times$   &-          &MIDI &1.5M\\
            \cmidrule(lr){1-5}
            MuseScore (ours)            &\checkmark &\checkmark &XML  &1.5M\\
            \bottomrule
    \end{tabular}
\end{table}

\begin{figure*}
    \footnotesize
    \centering
    \includegraphics[width=\linewidth]{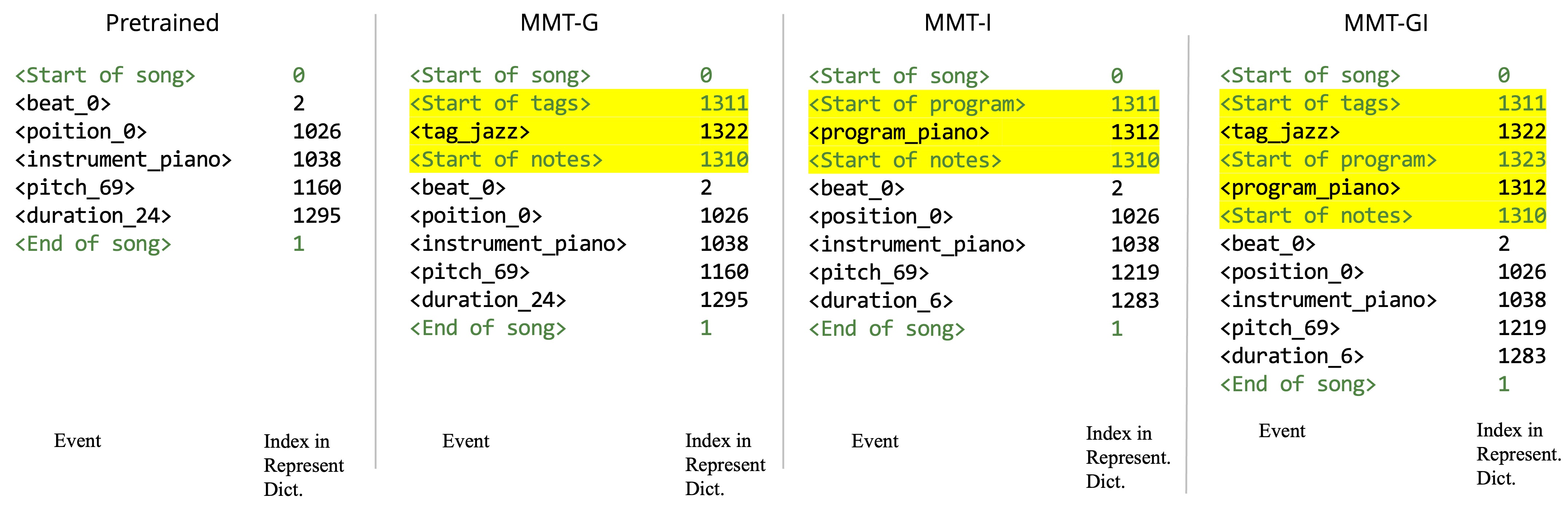}\\
    (a)\hspace{15em}(b)\hspace{15em}(c)\hspace{15em}(d)
    \captionsetup{font=footnotesize}
    \caption{An example of the baseline representation and our proposed representations---(a) pretrained, (b) genre conditioned generation used by MMT-G, (c) instrument conditioned generation used by MMT-I, and (d) genre-instrument conditioned generation used by MMT-GI. We highlight the newly added tokens in (b)--(d). The temporal resolution is 12 in this example.}
    \label{fig:representation}
\end{figure*}

%===============
\section{Method}
%===============

%-------------------------------
\subsection{Data Representation}
%-------------------------------
We represent a music piece as a one-dimensional array of integers using an event-based representation
adapted from REMI+ \cite{FIGARO} and MMT \cite{MMT}. REMI+ represents notes with six consecutive tokens encoding note position, pitch, velocity, duration, instrument and time-signature information. However, it cannot provide control over genre and instruments. MMT represents a sequence of six-dimension events, with each event $x_{i}$ encoded as a tuple of variables ($x^\mathit{type}$, $x^\mathit{beat}$, $x^\mathit{position}$, $x^\mathit{pitch}$, $x^\mathit{duration}$, $x^\mathit{instrument}$). Nevertheless, given that MMT processes the six output fields independently, it does not account for the potential interdependencies within these fields for a specific note. As a result, we adapt the REMI+ representation \cite{FIGARO} to provide control over genre and instruments, while preserving the expressiveness offered by REMI+. Similar to MMT \cite{MMT}, we decompose note-on events to beat and position to reduce the size of the vocabulary and to help the model learn the rhythmic structure of music. In addition, we did not include the ``tempo'' and ``chord'' events as such information is not generally available in our dataset.

Following REMI+ \cite{FIGARO}, we use \texttt{beat}, \texttt{position}, \texttt{pitch} and \texttt{duration} events for representing musical notes. The \texttt{beat} event $x^\mathit{beat}$ denotes the index of the beat that the note lies in. The value of position $x^\mathit{position}$ is determined by subtracting the product of the temporal resolution per beat and $x^\mathit{beat}$ from the actual onset of the note. Moreover, we introduce the \texttt{instrument} and \texttt{tag} events for specifying the instruments and tags. In addition to these data tokens, we have five special structural events:
\begin{itemize}
    \item \texttt{start-of-song} event indicates the onset of a song.
    \item \texttt{start-of-program} event marks the beginning of Instruments, followed by a list of Instrument events.
    \item \texttt{start-of-tags} event marks the beginning of tags, followed by a list of tag events.
    \item \texttt{start-of-notes} event marks the beginning of notes, followed by a list of note events.
    \item \texttt{end-of-song} event indicates the end of the song, after which the model will stop predicting new tokens.
\end{itemize}

To facilitate controllability in the model, we propose to add `control tokens' at the start of the data representation, as shown in \cref{fig:representation}. The control tokens include tag and program. This model capitalizes on the autoregressive nature of the transformer model, enabling the integration of these tokens during the inference process, subtly enhancing the model's controlability. 
Note that the \texttt{start-of-notes} token is used to mark the end of tag and instrument lists to prevent the model from continuing to generate instrument or tag events.
The pretrained model has acquired a significant amount of features from the dataset. We show three examples in \cref{fig:representation} and highlight the new tokens that we add.

\begin{table*}
    \footnotesize
    \centering
    \caption{Objective Evaluation Results.\\\normalfont(Mean values and 95\% confidence intervals are reported. A closer value to that of the ground truth is considered better.)}
    \label{tab:objective result}
    \begin{tabular}{lccccccc}
        \toprule
        & \multicolumn{2}{c}{Condition} & \multirow{2}{*}[-6pt]{\shortstack{Model\\size}} & \multirow{2}{*}[-5pt]{\shortstack{Training\\samples}} & \multirow{2}{*}[-6pt]{\shortstack{Pitch class \\entropy}} & \multirow{2}{*}[-5pt]{\shortstack{Scale consistency\\(\%)}} &\multirow{2}{*}[-5pt]{\shortstack{Groove consistency\\(\%)}}\\
        \cmidrule(lr){2-3}
        Model & Instrument & Genre\\
        \midrule
        Pretrained & $\times$ & $\times$ & 87.15M & 1.35M & 2.70 $\pm$ 0.08 & 95.92 $\pm$ 1.70 & 91.57 $\pm$ 2.10\\
        \cmidrule(lr){1-8}
        MMT-I & \checkmark & $\times$ & 87.27M & 739K & 2.78 $\pm$ 0.08 & 94.36 $\pm$ 1.82 & 92.86 $\pm$ 1.40 \\
        Ground truth & - & - & - & - & 2.48 $\pm$ 0.18 & 96.11 $\pm$ 1.36 & 92.58 $\pm$ 1.36\\
        \cmidrule(lr){1-8}
        MMT-G & $\times$ & \checkmark & 87.18M & 158K & 2.88 $\pm$ 0.06 & 92.72 $\pm$ 2.08 & 92.58 $\pm$ 1.20\\
        MMT-GI & \checkmark & \checkmark & 87.28M & 158K & 2.86 $\pm$ 0.07 & 94.75 $\pm$ 1.90 & 92.48 $\pm$ 1.82\\
        Ground truth & - & - & - & - & 2.61 $\pm$ 0.13 & 96.10 $\pm$ 1.36 & 92.55 $\pm$ 1.36\\
        \bottomrule
    \end{tabular}
\end{table*}

\begin{figure}
    \small
    \centering
    \includegraphics[width=.6\linewidth]{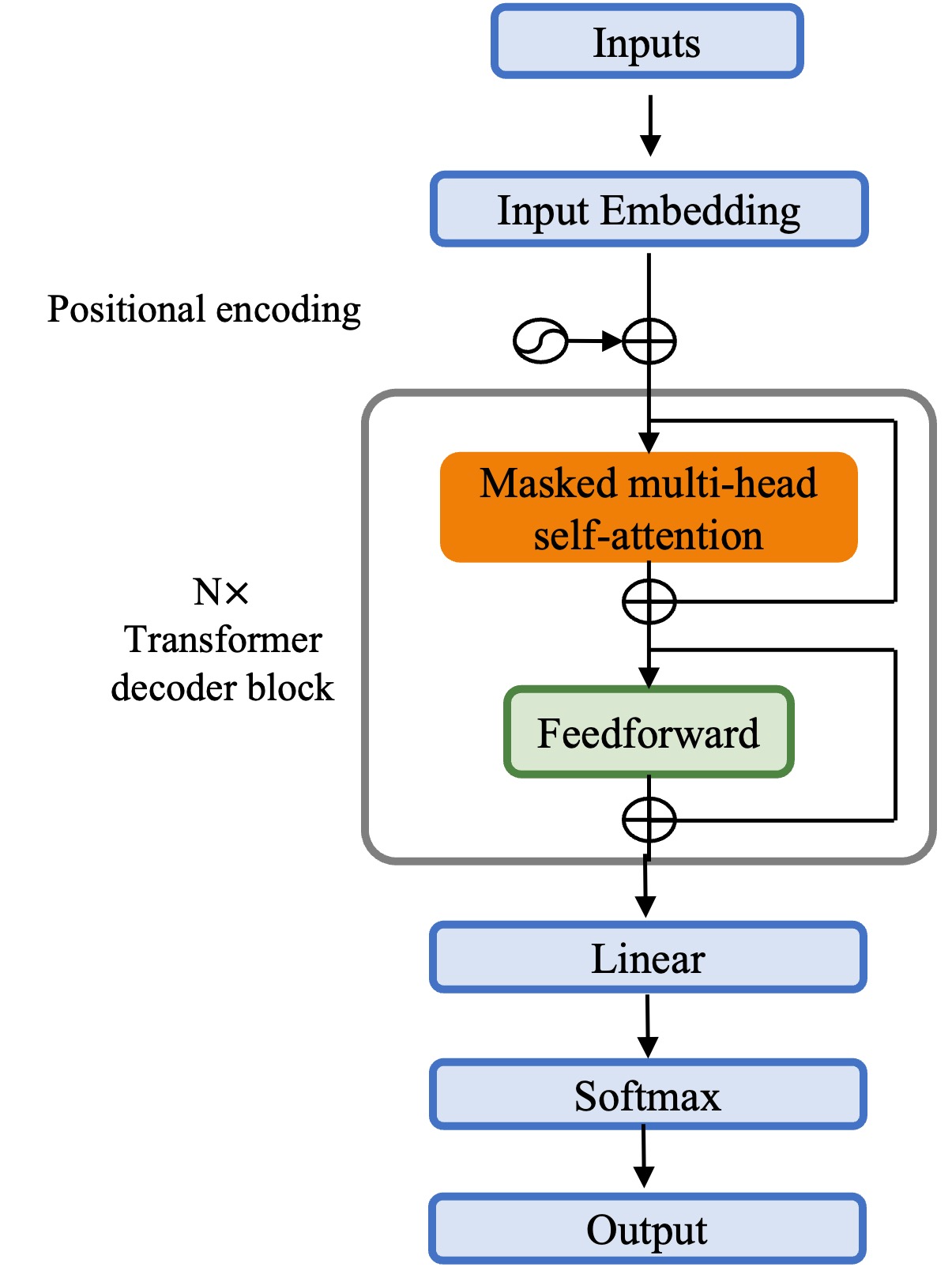}
    \caption{Illustration of the pretrained model structure}
    \label{fig:transformer}
\end{figure}
%------------------
\subsection{Models}
%------------------

We first train an unconditional model, as shown in \cref{fig:transformer} then fine-tune the unconditionally pretrained model. We initialize the embeddings for tokens that already exist in the pretrained model with the corresponding pretrained weights, while assigning random weights to any newly introduced tokens. In order to equip the model with high level controllability over genre and instruments, the model inputs are one-dimensional input sequences with prefixing conditions. We propose three variants of models---MMT-I, MMT-G and MMT-GI, which offer different controls as specified below.

\paragraph*{MMT-I for instrument conditioned generation.}
%--------------------------------------------------------

In this scenario, this model is provided with three event types: \texttt{start-of-song}, \texttt{start-of-program}, and \texttt{start-of-notes}, which serve to indicate the type of information that follows them. Then the model generates note sequences based on the predefined program lists. \texttt{end-of-notes} marks the end of the note sequences. 
See \cref{fig:representation}(b) for an example.

\paragraph*{MMT-G for genre conditioned generation.}
%---------------------------------------------------

This model is provided with three event types: \texttt{start-of-song}, \texttt{start-of-tags}, and \texttt{start-of-notes}, which serve to denote the type of information that follows them. Then the model generates notes sequence based on the predefined tag lists. \texttt{end-of-notes} marks the end of the note sequences.
See \cref{fig:representation} (c) for an example.

\paragraph*{MMT-GI for genre-instrument conditioned generation.} 
%---------------------------------------------------------------

This model is equipped with four event types: \texttt{start-of-song}, \texttt{start-of-tags}, \texttt{start-of-program}, and \texttt{start-of-notes}, which are used to identify the type of information that follows each event. We put the genre information ahead of the instrument list as we believe genre offers a higher-level control than instrumentation.\texttt{end-of-notes} marks the end of the note sequences.
See \cref{fig:representation} (d) for an example.

%==============================
\section{Experiments \& Results}
%===============================

%----------------------------------
\subsection{Implementation Details}
%----------------------------------

We use a model dimension of 768 and 12 attention heads for the transformer model. The maximum sequence length is 1024, and the maximum number of beats is 64. We use a temporal resolution of 12 time steps per quarter note for all the models. There are 87.15M, 87.27M, 87.18M and 87.28M parameters in the pretrained model, MMT-I, MMT-G and MMT-GI models, respectively. Different models are trained on different subsets based on the condition. We train the MMT-I model using the MuseScore-metadata dataset. We train the MMT-G and MMT-GI models with the MuseScore-genre dataset. We reserve 5\% of each dataset for validation and 5\% for test purpose for all the datasets. We validate the models every 10K steps. We pretrain the unconditional model on MuseScore-full for 1 million steps using a batch size of 32 on a NVIDIA RTX 2080 Ti GPU. The MMT-I, MMT-G and MMT-GI models are finetuned for 200K steps on a NVIDIA RTX A6000 GPU using a batch size of 16. For both the pretrained and finetuned models, we use the AdamW \cite{adamw} optimizer with $\beta_1 = 0.9$, $\beta_2 = 0.999$ and $\lambda = 0.01$. For the pretrained model, the learning rate starts with 0.0005. For the finetuned model, the learning rate starts with 0.0001. For both pretrained and finetuned models, we apply a linear learning rate decay schedule so that the learning rate decays to 10\% of its initial value in the first 100K steps.

%----------------------------------------
\subsection{Objective Evaluation Metrics}
%----------------------------------------

To evaluate our proposed models, we follow \cite{muspy} and compute the pitch class entropy, scale consistency and groove consistency:
\begin{itemize}
    \item \textit{Pitch class entropy} is determined by calculating the Shannon entropy of the normalized note pitch class histogram, with the exclusion of drum tracks.
    \item \textit{Scale consistency} is measured as the highest pitch-in-scale rate observed across all major and minor scales, also excluding drum tracks.
    \item \textit{Groove consistency} is assessed by computing the average Hamming distance between neighboring measures.
\end{itemize}
Our primary focus lies in discerning the distribution differences in these three criteria between genuine music and generated music.

%----------------------------------------
\subsection{Objective Evaluation Results}
%----------------------------------------

 We randomly generate 50 songs for each model and show in \cref{tab:objective result} the computed objective metrics. When evaluating the quality of generated music, we consider it better if the values are closer to those of the ground truth. When comparing MMT-G to MMT-GI, we do not observe a noticeable difference in terms of pitch class entropy, scale consistency and groove consistency. However, in terms of pitch class entropy, a significant statistical disparity is observed between MMT-G and the ground truth, and similarly, between MMT-GI with ground truth, indicating that there is still room for improvements for both models from a machine learning perspective. Moreover, in practice, we notice that the proposed models perform notably better on generating music for more common instruments (e.g., ``piano'') and genres (e.g., ``classical'') seen in the dataset (see \cref{fig:condition}). We encourage the readers to listen to the audio samples provided in the anonymous website.\cref{fn:demo}

\begin{table*}
    \footnotesize
    \centering
    \caption{Subjective Listening Test Results.\\\normalfont(Mean values and 95\% confidence intervals are reported.)}
    \label{tab:subjective test}
    \begin{tabular}{lccccccccc}
        \toprule
        & \multicolumn{2}{c}{Condition} & \multirow{2}{*}[-6pt]{\shortstack{Model\\size}} & \multirow{2}{*}[-5pt]{\shortstack{Training\\samples}} & & & & \multirow{2}{*}[-6pt]{\shortstack{Condition\\adherence}}\\
        \cmidrule(lr){2-3}
        Model & Instrument & Genre & & & Coherence & Harmony & Arrangement & & Overall\\
        \midrule
        Pretrained & $\times$ & $\times$ & 87.15M & 1.35M & 3.44 $\pm$ 0.51 & 3.28 $\pm$ 0.46 & 3.16 $\pm$ 0.48 & - & 3.12 $\pm$ 0.45\\ 
        \cmidrule(lr){1-10}
        MMT-I & \checkmark & $\times$ & 87.27M & 739K & 3.62 $\pm$ 0.49 & 3.22 $\pm$ 0.56 & 3.40 $\pm$ 0.48 & 3.15 $\pm$ 0.53 & 3.73 $\pm$ 0.55\\
        \cmidrule(lr){1-10}
        MMT-G & $\times$ & \checkmark & 87.18M & 158K & 3.83 $\pm$ 0.43 & 3.73 $\pm$ 0.46 & 3.66 $\pm$ 0.39 & 3.43 $\pm$ 0.47 & 3.37 $\pm$ 0.55 \\
        MMT-GI & \checkmark & \checkmark & 87.28M & 158K & 3.90 $\pm$ 0.48 & 3.80 $\pm$ 0.57 & 3.56 $\pm$ 0.62 & 3.60 $\pm$ 0.57 & 3.42 $\pm$ 0.55\\
        \bottomrule
    \end{tabular}
\end{table*}

%-------------------------------------
\subsection{Subjective Listening Test}
%-------------------------------------

Within each music generation task, we seek to understand specific attributes of the produced tunes. To this end, we conduct a listening test where each participant is instructed to assess 5 songs for every given condition. Out of the 11 participants in our study, 10 people have experience in playing instruments, with one being a professional musician.

\paragraph*{Music quality.}
%--------------------------

First, we assess the quality of the generated music in terms of coherence, harmoniousness and arrangement. The participants are instructed to answer the following questions in a Likert scale of 1 to 5.
\begin{itemize}
    \item \textit{Coherence}: Is it temporally coherent? Is the rhythm steady? Are there many out-of-context notes?
    \item \textit{Harmoniousness}: Is it harmonious?
    \item \textit{Arrangement}: Are the instruments used reasonably? Are the instruments arranged properly?
\end{itemize}

\paragraph*{Condition Adherence.}
%----------------------------

We evaluate whether the generated music aligns with the predefined tags or Instruments. This is measured by a score for how relevant the generated samples to the desired genre/instrument list. We ask participants the following question: ``\textit{How relevant the generate samples are to the desired genre and instruments?}''

\paragraph*{Overall Performance.}
%----------------------------

We also ask the participants how they generally feel about the generated music with the following questions ``\textit{Considering all criteria above, how much do you think it is a good song that can meet the condition while maintain coherence, harmony and arrangement?  Is it pleasant to listen to? How close is it to real music? Does it match the condition we set?}''

%-------------------------------------
\subsection{Subjective Listening Test Results}
%-------------------------------------

We report in \cref{tab:subjective test} the mean opinion scores (MOS) and the 95\% confidence intervals assuming a Gaussian distribution. We can see our models can generate qualified music. We do not observe a significant difference between MMT-G and MMT-GI in terms of coherence, harmony and arrangement. Moreover, we see that MMT-I significantly outperforms the pretrained baseline model in terms of overall quality. Interestingly, while MMT-GI has more conditions to follow, it achieves a high score in terms of condition adherence as compared to MMT-I and MMT-G, which have only one condition to follow.

%===================================
\section{Discussions \& Future Work}
%===================================

In the subjective test, we observe an intriguing pattern: music generated with genre-instrument conditioning more closely adheres to the predefined conditions. This might stem from the inherent entanglement of genre and instrumentation information.

For future work, we want to explore the following three directions. First, the metadata of the MuseScore dataset contains other valuable attributes such as composer, key signature, time signature, popularity and rating information, which can serve as powerful conditioning signals. While we consider only genre and instrument information in this work, we would like to extend our model to include additional controls to further improve its high-level controllability.

Second, we note that the music our model produces, especially when the genre is set to classical, does not exhibit a quality comparable to that of MMT \cite{MMT}. This observation is from an informal, internal side-by-side listening comparison between samples from our model and those from MMT\cite{MMT}. A potential reason is the presence of noisy and low-quality entries in MuseScore. To enhance the quality of our dataset, we intend to remove low-quality songs such as practice sessions and incomplete songs as well as those without enough metadata. Furthermore, we plan to integrate other high-quality datasets such as POP909 \cite{pop909} and SOD \cite{sod}.

Third, we find that most of the generated music leans heavily towards certain types of music, notably classical and piano compositions. As illustrated in \cref{fig:condition}, MuseScore is an imbalanced dataset that biases towards classical and piano compositions. The long-tailed nature of the dataset poses challenges to generating music of less common categories. Therefore, it is crucial to adopt a smarter sampling strategy that compensates underrepresented genres and instruments.

We plan to downsample common genres and instruments, and adapt advanced sampling and weighting method from other field \cite{noiselabel}. Additionally, it might be helpful to include music pieces in underrepresented genres from other genre-specific datasets.

%===================
\section{Conclusion}
%===================

In this work, we first pretrained a large and unconditional transformer model. Then we finetuned the pretrained model with new control tokens to equip the model with genre and instrument controls. We have shown through our experiments that the proposed model can successfully produce music of the specified genre and instruments. The model also outperforms the pretrained baseline model in the subjective listening test. Our study has shown that the ``pretraining-and-finetuning'' paradigm can be a promising direction worth further exploration for music generation.

%=========================
\section{Acknowledgements}
%=========================

Hao-Wen thanks Taiwan Ministry of Education for supporting his PhD study. This project has received funding from the European Research Council (ERC REACH) under the European Union’s Horizon 2020 research and innovation programme (Grant agreement \#883313).

\end{document}